\documentstyle[prl,floats,aps,epsf]{revtex}

\begin{document}
\twocolumn   

\title{Information-Theoretic Description of No-go Theorem \\
of a Bit Commitment}
\author{Yoshihiro Nambu\cite{byline1} and Yoshie Chiba-Kohno\cite{byline2}}
\address{Fundamental Research Laboratories, System Devices and Fundamental\\
Research, NEC Corporation, 34 Miyukigaoka, Tsukuba, Ibaraki 305-8501, Japan}
\date{\today }
\maketitle

\begin{abstract}
We give a comprehensive and constructive proof of the no-go theorem of a bit
commitment given by Mayers, Lo, and Chau from the viewpoint of quantum
information theory. It is shown that there is a trade-off relation between
information acquired by Bob during the commitment phase and the ability to
change a commit bit by Alice during the opening phase. It is clarified that
a protocol that is unbiased to both Alice and Bob cannot be, at the same
time, secure against both parties. Fundamental physical constraints that
govern this no-go theorem are also discussed.
\end{abstract}


\narrowtext    


\preprint{20001115}

\section{Introduction}

\label{SEC1}

In brief, a bit commitment (BC) is the following task that is executed in
two steps ((a) and (b) below) by two mistrustful parties, a sender, Alice,
and a receiver, Bob.\newline
(a) Commit phase (C-phase): Alice chooses a bit ($b=0$ or $1$) and commits
it to Bob. That is, she gives Bob a piece of evidence that she has a bit $b$
in mind and that she cannot change it (in this case, the commitment is said
to be binding). Bob cannot learn the value of the committed bit from that
evidence until Alice reveals further information (in this case, the
commitment is said to be concealing).\newline
(b) Opening phase (O-phase): At a later time, Alice opens the commitment.
That is, she tells Bob the value of $b$ and convinces him that it is indeed
the genuine bit that she chose during the C-phase. If Alice changes the
value, it can be discovered by Bob.

A BC is an important cryptographic primitive with many applications in more
sophisticated tasks and is of great theoretical and practical interest.
Current classical BC protocols are proven secure by invoking some unproven
computational assumption; that is, complexity of some kind of mathematical
problems such as the hardness of factoring large integers. After the
invention of the quantum computing algorithm that makes the computational
assumption totally invalid, it has been brought to many researchers'
attention whether there exists a BC protocol that is guaranteed secure
solely by physical principles. In recent years, Mayers, Lo, and Chau have
proven that an unconditionally secure BC is impossible (no-go theorem for a
BC) under the standard nonrelativistic assumption.\cite{Mayers,Lo} However,
although their discussions are quite correct, their proofs are a bit formal
and nonconstructive. It is not yet clear what prevents us from implementing
the unconditionally secure BC protocol. In this paper, we give a
constructive proof of the no-go theorem for a BC that would make things more
transparent and convincing from the viewpoint of quantum information theory.
We clarify why quantum mechanics does not help a quantum BC protocol to
achieve more than a classical one does.

\section{Model and formulation of the bit commitment protocol}

\label{SEC2}

First, let us consider an honest protocol. The most important point
concerning the BC protocol is that Alice needs to unveil a value of $b$ in
the O-phase consistently with information transmitted in the C-phase. From
the information-theoretic point of view, this implies that one bit of
classical information should be transmitted from Alice to Bob at the end of
the protocol. Therefore, when we set $I_{c}$ and $I_{o}$ as the amounts of
information in a bit transmitted in the C-phase and O-phase, the following
identity holds: 
\begin{equation}
I_{c}+I_{o}=1.  \label{eq1}
\end{equation}%
Namely, the BC protocol is essentially a split transmission of one-bit
information in two temporally separated steps: one in the C-phase and the
other in the O-phase. Only a fraction of the bit information needs to be
transmitted in each step. Noting this fact, we can formulate the quantum BC
protocols reported so far\cite{Bennett,Brassard1,Brassard2,Hardy} as follows.

In order to demand unconditional security, Alice reveals to Bob quantum
information as a piece of evidence of her commitment by transmitting a
system, such as a photon or an electron, in the C-phase. In the O-phase, she
reveals to Bob classical information which consists of the value of $b$ and
the measurement basis on the system. Finally, Alice and Bob test the
consistency between the reported value of $b$ and the measurement results of
the system.

According to the quantum description of the protocol involving classical
communication suggested by Tal Mor, let subsystem $B$ (Bob's system) be the
system with arbitrary dimensional state space $H_{B}$ that carries quantum
information in the C-phase and subsystem $A$ (Alice's system) be the system
with arbitrary dimensional state space $H_{A}$ that carries classical
information in the O-phase.\cite{TalMor} Let $\chi _{b}^{AB}$ be the genuine
states of the joint system $AB$ to be prepared by Alice according to her
choice of $b$. Then, Eq. (\ref{eq1}) is equivalent to the condition that $%
\chi _{0}^{AB}$ and $\chi _{1}^{AB}$ are orthogonal in the joint Hilbert
space, $H_{AB}=H_{A}\otimes H_{B}$; i.e., 
\begin{equation}
\chi _{0}^{AB}\chi _{1}^{AB}=\chi _{1}^{AB}\chi _{0}^{AB}=0.  \label{eq2}
\end{equation}%
Here, to avoid confusion throughout this paper, we use superscripts to
denote the appropriate state space for a state or an operator. According to
the protocol, by transmitting a subsystem $B$ to Bob, Alice reveals, in
general, nonorthogonal marginal states 
\begin{equation}
\rho _{b}^{B}=Tr_{A}\chi _{b}^{AB}  \label{eq3}
\end{equation}%
($b=0$ or $1$) in the C-phase. It is proven in Appendix A that from
condition (\ref{eq2}), we can always find two mutually orthogonal
purifications $\left| \psi _{b}^{AB}\right\rangle $ of $\rho _{b}^{B}$ in
the orthogonal subspace in which the support of the state $\chi _{b}^{AB}$
lies. Any set of two orthonormal states in $H_{AB}$ should be represented by
two orthonormal states in the two-dimensional subspace $M$ spanned by $%
\left\{ \left| 0^{AB}\right\rangle ,\left| 1^{AB}\right\rangle \right\} $ in 
$H_{AB}$. Furthermore, any such subspace $M$ can be defined by a set of
two orthonormal states in $H_{AB}$: 
\begin{eqnarray}
\left| 0^{AB}\right\rangle  &=&\sum \alpha \left| a^{\prime A}\right\rangle
\left| a^{B}\right\rangle ,  \nonumber \\
\left| 1^{AB}\right\rangle  &=&\sum \beta \left| b^{\prime A}\right\rangle
\left| b^{B}\right\rangle ,  \label{eq5}
\end{eqnarray}%
with an appropriate choice of two sets of orthonormal states $\left\{ \left|
a^{\prime A}\right\rangle \left| a^{B}\right\rangle \right\} $ and $\left\{
\left| b^{\prime A}\right\rangle \left| b^{B}\right\rangle \right\} $ and
coefficients $\alpha $ and $\beta $, \bigskip where $\left\{ \left\{ \left|
a^{\prime A}\right\rangle \right\} ,\left\{ \left| b^{\prime A}\right\rangle
\right\} \right\} $ $(\left\{ \left\{ \left| a^{B}\right\rangle \right\}
,\left\{ \left| b^{B}\right\rangle \right\} \right\} )$ makes up a Schmidt
basis for $H_{A}$ $(H_{B})$. Therefore, given a set of the mutually
orthogonal purifications $\left\{ \left| \psi _{0}^{AB}\right\rangle ,\left|
\psi _{1}^{AB}\right\rangle \right\} $ in $H_{AB}$, we can always find the
following form of Schmidt decomposition\cite{Hughston,Ekert,Barnett}, 

\begin{equation}
\left\{ 
\begin{array}{l}
\left| \psi _{0}^{AB}\right\rangle =\cos \theta \left| 0^{AB}\right\rangle
+\sin \theta \left| 1^{AB}\right\rangle , \\[2mm]
\left| \psi _{1}^{AB}\right\rangle =-\sin \theta \left| 0^{AB}\right\rangle
+\cos \theta \left| 1^{AB}\right\rangle ,%
\end{array}%
\right.   \label{eq4}
\end{equation}%
by choosing appropriate Schmidt coefficients and absorbing any phase factors
in the definition of the bases. As a result of Eqs. (\ref{eq5}) and (\ref%
{eq4}), the marginal states $\rho _{0}^{B}$ and $\rho _{1}^{B}$ are
commutable and diagonalized simultaneously by Schmidt basis as 
\begin{eqnarray}
\rho _{0}^{B} &=&Tr_{A}\left| \psi _{0}^{AB}\right\rangle \left\langle \psi
_{0}^{AB}\right| =\cos ^{2}\theta \hat{\rho}_{0}^{B}+\sin ^{2}\theta \hat{%
\rho}_{1}^{B},  \nonumber \\
\rho _{1}^{B} &=&Tr_{A}\left| \psi _{1}^{AB}\right\rangle \left\langle \psi
_{1}^{AB}\right| =\sin ^{2}\theta \hat{\rho}_{0}^{B}+\cos ^{2}\theta \hat{%
\rho}_{1}^{B},  \label{eq6}
\end{eqnarray}%
where $Tr_{A}$ denotes a partial trace over subsystem $A$, and two states 
\begin{eqnarray}
\hat{\rho}_{0}^{B} &=&Tr_{A}\left| 0^{AB}\right\rangle \left\langle
0^{AB}\right| =\sum \alpha ^{2}\left| a^{B}\right\rangle \left\langle
a^{B}\right| ,  \nonumber \\
\hat{\rho}_{1}^{B} &=&Tr_{A}\left| 1^{AB}\right\rangle \left\langle
1^{AB}\right| =\sum \beta ^{2}\left| b^{B}\right\rangle \left\langle
b^{B}\right| ,  \label{eq7}
\end{eqnarray}%
are orthogonal on $H_{B}$; i.e., $\hat{\rho}_{0}^{B}\hat{\rho}_{1}^{B}=\hat{%
\rho}_{1}^{B}\hat{\rho}_{0}^{B}=0$. The forms of $\hat{\rho}_{0}^{B}$ and $%
\hat{\rho}_{1}^{B}$ can be freely chosen in the protocol and various complex
forms have been proposed to prevent cheating of both parties, but the
concrete forms are irrelevant to the subject in the following discussion.

\section{Cheating strategies}

\label{SEC3}

According to the model given in Sec \ref{SEC2}, we will evaluate the
performance of Alice's and Bob's cheating.

\subsection{Bob's cheating}

The purpose of Bob's cheating is to obtain as much information as possible
about $b$ during the C-phase from the marginal states $\rho _{b}^{B}$. In
the following, the amount of available information about $b$ for Bob during
the C-phase is evaluated as a measure of his cheating performance.

From the protocol agreed by Alice and Bob, the states $\chi _{b}^{AB}$ to be
prepared by Alice are known to them. Therefore, Bob can calculate the
Schmidt bases, $\left\{ \left| a^{B}\right\rangle \right\} $ and $\left\{
\left| b^{B}\right\rangle \right\} $, that diagonalize the marginal states $%
\rho _{0}^{B}$ and $\rho _{1}^{B}$ beforehand. Bob can perform an optimal
measurement for distinguishing $\rho _{0}^{B}$ and $\rho _{1}^{B}$ by making
use of the orthogonality between $\hat{\rho}_{0}^{B}$ and $\hat{\rho}%
_{1}^{B} $. To describe a measure of the available information about $b$ for
Bob from $\rho _{b}^{B}$, consider the fidelity between $\rho _{0}^{B}$ and $%
\rho _{1}^{B}$. \cite{Jozsa,Fuchs} It is given by 
\begin{equation}
F(\rho _{0}^{B},\rho _{1}^{B})=Tr_{B}\sqrt{\left( \rho _{1}^{B\text{ }%
}\right) ^{1/2}\rho _{0}^{B}\left( \rho _{1}^{B\text{ }}\right) ^{1/2}}.
\label{eq8}
\end{equation}%
Noting that $\hat{\rho}_{0}^{B}$ and $\hat{\rho}_{1}^{B}$ are orthogonal, we
can calculate $\left( \rho _{1}^{B\text{ }}\right) ^{1/2}$ from Eq. (\ref%
{eq6}) as 
\begin{equation}
\left( \rho _{1}^{B\text{ }}\right) ^{1/2}=\left| \sin \theta \right| \left( 
\hat{\rho}_{0}^{B}\right) ^{1/2}+\left| \cos \theta \right| \left( \hat{\rho}%
_{1}^{B}\right) ^{1/2}  \label{eq9}
\end{equation}%
in the representation in which $\hat{\rho}_{0}^{B}$ and $\hat{\rho}_{1}^{B}$
are diagonal. Therefore, Eq. (\ref{eq6}) gives%
\begin{equation}
F(\rho _{0}^{B},\rho _{1}^{B})=\frac{1}{2}\left| \sin 2\theta \right|
Tr_{B}\left( \hat{\rho}_{0}^{B}+\hat{\rho}_{1}^{B}\right) =\left| \sin
2\theta \right| .  \label{eq10}
\end{equation}

The smaller the fidelity is, the more Bob can distinguish between $\rho
_{0}^{B}$ and $\rho _{1}^{B}$ correctly; therefore, he can gain more
information about $b$. To confirm this, we consider the quantum error
probability which gives the lower limit of error rate for distinguishing $%
\rho _{0}^{B}$ and $\rho _{1}^{B}$.\cite{Fuchs,Fuchs2,Helstrom} It is given
as 
\begin{equation}
P_{err}^{Bob}(\rho _{0}^{B},\rho _{1}^{B})=\frac{1}{2}-\frac{1}{4}%
Tr_{B}\left| \rho _{0}^{B}-\rho _{1}^{B}\right| .  \label{eq11}
\end{equation}%
Noting again that $\hat{\rho}_{0}^{B}$ and $\hat{\rho}_{1}^{B}$ are
orthogonal, it follows from Eq. (\ref{eq6}) that 
\begin{equation}
\rho _{0}^{B}-\rho _{1}^{B}=\cos 2\theta \left( \hat{\rho}_{0}^{B}-\hat{\rho}%
_{1}^{B}\right)  \label{eq12}
\end{equation}%
in the representation in which $\hat{\rho}_{0}^{B}$ and $\hat{\rho}_{1}^{B}$
are diagonal. Thus, it follows that 
\begin{equation}
P_{err}^{Bob}(\rho _{0}^{B},\rho _{1}^{B})=\frac{1-\left| \cos 2\theta
\right| }{2}=\frac{1-\sqrt{1-\left( F(\rho _{0}^{B},\rho _{1}^{B})\right)
^{2}}}{2}.  \label{eq13}
\end{equation}%
Let us now introduce distinguishability between $\rho _{0}^{B}$ and $\rho
_{1}^{B}$ as 
\begin{equation}
D(\rho _{0}^{B},\rho _{1}^{B})=P_{cor}^{Bob}-P_{err}^{Bob}=\frac{1}{2}%
Tr_{B}\left| \rho _{0}^{B}-\rho _{1}^{B}\right| .  \label{eq14}
\end{equation}%
Then, the larger the distinguishability is, the more Bob can distinguish
between $\rho _{0}^{B}$ and $\rho _{1}^{B}$ correctly. Thus, the
distinguishability gives a measure of the available information about $b$
for Bob from $\rho _{b}^{B}$. It is easily seen that $F(\rho _{0}^{B},\rho
_{1}^{B})$ and $D(\rho _{0}^{B},\rho _{1}^{B})$ satisfy 
\begin{equation}
\left( F(\rho _{0}^{B},\rho _{1}^{B})\right) ^{2}+\left( D(\rho
_{0}^{B},\rho _{1}^{B})\right) ^{2}=1.  \label{eq15}
\end{equation}%
Therefore, there is a trade-off relationship between $F(\rho _{0}^{B},\rho
_{1}^{B})$ and $D(\rho _{0}^{B},\rho _{1}^{B})$; that is, the smaller $%
F(\rho _{0}^{B},\rho _{1}^{B})$ is, the larger $D(\rho _{0}^{B},\rho
_{1}^{B})$ is and the more correctly Bob can distinguish $\rho _{0}^{B}$ and 
$\rho _{1}^{B}$, and vice versa.

Let us turn to the information-theoretic measure of available information
for Bob. Mutual information between the value of genuine $b$ and the value
of $b$ that is judged from the measurement of $\rho _{0}^{B}$ and $\rho
_{1}^{B}$ is an appropriate measure from the viewpoint of information
theory. When Alice chooses the value of commit bit $b$ between 0 and 1 with
equiprobability, this measure depends only on $\rho _{0}^{B}$ and $\rho
_{1}^{B}$ and is given by 
\begin{equation}
I^{Bob}(\rho _{0}^{B},\rho _{1}^{B})=1-H(P_{err}^{Bob}(\rho _{0}^{B},\rho
_{1}^{B})),  \label{eq16}
\end{equation}%
where $H(p)=-p\log _{2}p-(1-p)\log _{2}(1-p)$ is an entropy function (in
bit).

\subsection{Alice's cheating}

The purpose of Alice's cheating is to unveil her commit bit $b$ at her will
in the O-phase while ensuring unveiled $b$ does not conflict with Bob's
measurement of his subsystem $B$ revealed by her during the C-phase. From
the agreed protocol, Alice can calculate the purification given in Eqs. (\ref%
{eq4}) and (\ref{eq5}) beforehand.

In the following, her ability to change the commit bit is evaluated for two
known cheating strategies as a measure of her cheating performance.

\subsubsection{Mayer's strategy}

This is a strategy which was first proposed by Mayers.\cite{Mayers} Alice
honestly reveals either $\rho _{0}^{B}$ or $\rho _{1}^{B}$ in the C-phase by
transmitting subsystem $B$ of the joint system $AB$ prepared in the
arbitrary purification associated to either $\rho _{0}^{B}$ or $\rho _{1}^{B}
$. In the O-phase, by a local unitary operation on subsystem $A$ in her
hand, she can change the joint state into any purification $\left| \bar{\psi}%
_{\bar{b}}^{AB}\right\rangle $\ of her chosen $\rho _{b}^{B}$ that satisfies 
\begin{equation}
\rho _{b}^{B}=Tr_{A}\left| \bar{\psi}_{\bar{b}}^{AB}\right\rangle
\left\langle \bar{\psi}_{\bar{b}}^{AB}\right|   \label{eq17}
\end{equation}%
and 
\begin{equation}
0\leq \left\langle \psi _{b}^{AB}|\bar{\psi}_{\bar{b}}^{AB}\right\rangle
\leq F(\rho _{0}^{B},\rho _{1}^{B}),  \label{eq18}
\end{equation}%
where $\bar{b}=b\oplus 1$.\cite{Hughston,Jozsa} Then, according to her
necessity, she changes the joint state into the fake states, 
\begin{eqnarray}
\left| \bar{\psi}_{1}^{AB}\right\rangle  &=&-\cos \theta \left|
0^{AB}\right\rangle +\sin \theta \left| 1^{AB}\right\rangle ,  \nonumber \\
\left| \bar{\psi}_{0}^{AB}\right\rangle  &=&\sin \theta \left|
0^{AB}\right\rangle +\cos \theta \left| 1^{AB}\right\rangle .  \label{eq19}
\end{eqnarray}%
Here, the state $\left| \bar{\psi}_{b}^{AB}\right\rangle $ saturates the
upper bound of $\left\langle \psi _{b}^{AB}|\bar{\psi}_{\bar{b}%
}^{AB}\right\rangle $ in Eq. (\ref{eq18}) and is most parallel to the state $%
\left| \psi _{b}^{AB}\right\rangle $.\cite{Jozsa} She tells Bob the basis to
be used for his measurement on subsystem $B$ that is found from her
projection measurement of subsystem $A$ by an appropriate basis $\left\{
\left| e_{j}^{A}\right\rangle \right\} $. Bob performs projection
measurement on his subsystem $B$ according to her instruction and checks her
commitment from the consistency between the value of $b$ that is unveiled by
Alice in the O-phase and his measurement results.

The fake state $\left| \bar{\psi}_{\bar{b}}^{AB}\right\rangle $ given in Eq.
(\ref{eq19}) is optimal in Mayer's strategy. To confirm this, we consider
the probability $P_{err}^{Alice}$\ that Alice causes and Bob finds an
inconsistency between the unveiled value of $b$ and Bob's measured data. It
is proven in Appendix B that $P_{err}^{Alice}$ is zero when Alice prepares
the genuine state $\chi _{b}^{AB}$ or the purification $\left| \psi
_{b}^{AB}\right\rangle $, and 
\begin{equation}
P_{err}^{Alice}\geq 1-\left| \left\langle \psi _{b}^{AB}|\bar{\psi}_{\bar{b}%
}^{AB}\right\rangle \right| ^{2}  \label{eq20}
\end{equation}%
when she prepares the fake state $\left| \bar{\psi}_{\bar{b}%
}^{AB}\right\rangle $, where equality holds if and only if $\left| \bar{\psi}%
_{\bar{b}}^{AB}\right\rangle $ lies in the subspace $M$ spanned by a set of
orthonormal states $\left\{ \left| 0^{AB}\right\rangle ,\left|
1^{AB}\right\rangle \right\} $. Applying Eq. (\ref{eq18}) to Eq. (\ref{eq20}%
) yields 
\begin{equation}
P_{err}^{Alice}\geq 1-\left( F(\rho _{0}^{B},\rho _{1}^{B})\right) ^{2}=P_{M%
\text{ }err}^{Alice}(\rho _{0}^{B},\rho _{1}^{B}).  \label{eq21}
\end{equation}%
The state $\left| \bar{\psi}_{\bar{b}}^{AB}\right\rangle $ in Eq. (\ref{eq19}%
) yields the lower bound $P_{M\text{ }err}^{Alice}(\rho _{0}^{B},\rho
_{1}^{B})$ for the probability $P_{err}^{Alice}$ that depends only on $\rho
_{0}^{B}$ and $\rho _{1}^{B}$. Therefore, the fake states in Eq. (\ref{eq19}%
) give the least possibility of disclosing her cheating to Bob and they are
optimal for this strategy. The lower limit $P_{M\text{ }err}^{Alice}(\rho
_{0}^{B},\rho _{1}^{B})$ is a convenient measure of Alice's ability to
change her commitment in the O-phase.

It should be noted that Mayer's strategy is asymmetric with respect to the
value of $b$ that Alice unveils in the O-phase. For example, consider Alice
reveals $\rho _{0}^{B}$ in the C-phase. Then, if she unveils $b=0$ honestly
in the O-phase, Bob's measured data on his subsystem $B$ is perfectly
consistent with her disclosure and $P_{err}^{Alice}$ is zero. Conversely, if
she wants to unveil $b=1$, she can cheat Bob successfully with the
probability 
\begin{equation}
P_{M\text{ }err}^{Alice}(\rho _{0}^{B},\rho _{1}^{B})=\cos ^{2}2\theta 
\label{eq22}
\end{equation}%
by preparing the fake state $\left| \bar{\psi}_{1}^{AB}\right\rangle $.
Here, Eqs. (\ref{eq10}) and (\ref{eq21}) are used to derive Eq. (\ref{eq22}%
). Thus, the lower limit $P_{M\text{ }err}^{Alice}(\rho _{0}^{B},\rho
_{1}^{B})$ depends on $b$ that Alice unveils in the O-phase. It should be
further noted that $P_{M\text{ }err}^{Alice}(\rho _{0}^{B},\rho _{1}^{B})>1/2
$ if $F(\rho _{0}^{B},\rho _{1}^{B})<1/\sqrt{2}$. This means that Mayer's
strategy can be applicable only when $F(\rho _{0}^{B},\rho _{1}^{B})=\left|
\sin 2\theta \right| \geq 1/\sqrt{2}$.

Now let us turn to the information-theoretic measure of Alice's cheating
performance. Let $I_{M}^{Alice}$ be mutual information between the value of $%
b$ that Alice unveils and the value of $b$ that Bob judged from his
measurement on his subsystem $B$ in the O-phase. Taking into account the
asymmetry noted in the previous paragraph, we get the upper bound of $%
I_{M}^{Alice}$ as a function only of $\rho _{0}^{B}$ and $\rho _{1}^{B}$ as
follows: 
\begin{equation}
I_{M}^{Alice}(\rho _{0}^{B},\rho _{1}^{B})=\frac{1}{2}+\frac{1}{2}\left\{
1-H(P_{M\text{ }err}^{Alice}(\rho _{0}^{B},\rho _{1}^{B}))\right\} .
\label{eq23}
\end{equation}%
Here, $I_{M}^{Alice}(\rho _{0}^{B},\rho _{1}^{B})$ is considered to be a
good information-theoretic measure of Alice's ability to change her commit
bit for this strategy.

\subsubsection{Hardy-Kent's strategy}

This is a strategy which was first given by Koashi and Imoto in the context
of quantum key distribution\cite{Koashi}, but later applied to the BC
protocol by Hardy and Kent.\cite{Hardy} According to this strategy, Alice
reveals $\bar{\rho}^{B}=\left( \rho _{0}^{B}+\rho _{1}^{B}\right) /2$ in the
C-phase by transmitting the subsystem $B$ of the joint system $AB$ prepared
in the arbitrary purification of $\bar{\rho}^{B}$. When she unveils her
commitment in the O-phase, she can change the joint state into any
purification $\left| \bar{\psi}_{b}^{AB}\right\rangle $ of $\bar{\rho}^{B}$
satisfying 
\begin{equation}
\bar{\rho}^{B}=Tr_{A}\left| \bar{\psi}_{b}^{AB}\right\rangle \left\langle 
\bar{\psi}_{b}^{AB}\right|  \label{eq24}
\end{equation}%
and 
\begin{equation}
0\leq \left\langle \psi _{b}^{AB}|\bar{\psi}_{b}^{AB}\right\rangle \leq
F(\rho _{b}^{B},\bar{\rho}^{B})  \label{eq25}
\end{equation}%
by performing a local unitary operation on her subsystem $A$. Then,
according to her choice of $b$, she changes the joint state into the fake
state, for example, so that when $0\leq \theta \leq \pi /2$, 
\begin{eqnarray}
\left| \bar{\psi}_{0}^{AB}\right\rangle &=&\left( \left| 0^{AB}\right\rangle
+\left| 1^{AB}\right\rangle \right) /\sqrt{2},  \nonumber \\
\left| \bar{\psi}_{1}^{AB}\right\rangle &=&-\left( \left|
0^{AB}\right\rangle -\left| 1^{AB}\right\rangle \right) /\sqrt{2}.
\label{eq26}
\end{eqnarray}%
Here, $\left| \bar{\psi}_{b}^{AB}\right\rangle $ saturates the upper bound
of $\left\langle \psi _{b}^{AB}|\bar{\psi}_{b}^{AB}\right\rangle $ in Eq. (%
\ref{eq25}) and is the most parallel to the state $\left| \psi
_{b}^{AB}\right\rangle $.\cite{Jozsa} She tells Bob the basis to be used for
his measurement on subsystem $B$ that is found from her projection
measurement on her subsystem $A$ by an appropriate basis $\left\{ \left|
e_{j}^{A}\right\rangle \right\} $. Bob performs projection measurement on
his system according to her instruction and checks her commitment from the
consistency between the value of $b$ that is unveiled by Alice in the
O-phase and his measurement results.

It can also be proven from Appendix B that the lower bound $P_{HK\text{ }%
err}^{Alice}(\rho _{0}^{B},\rho _{1}^{B})$ of the probability $%
P_{err}^{Alice}$\ in this strategy depends only on $\rho _{0}^{B}$ and $\rho
_{1}^{B}$, and it is given by 
\begin{equation}
P_{HK\text{ }err}^{Alice}(\rho _{0}^{B},\rho _{1}^{B})=1-\left( F(\rho
_{b}^{B},\bar{\rho}^{B})\right) ^{2}=\frac{1-F(\rho _{0}^{B},\rho _{1}^{B})}{%
2}.  \label{eq27}
\end{equation}%
The states $\left| \bar{\psi}_{b}^{AB}\right\rangle $ in Eq. (\ref{eq26})
yield the lower bound $P_{HK\text{ }err}^{Alice}(\rho _{0}^{B},\rho
_{1}^{B}) $. Therefore, they are optimal for this strategy. The lower limit $%
P_{HK\text{ }err}^{Alice}(\rho _{0}^{B},\rho _{1}^{B})$ gives a convenient
measure of Alice's ability to change her commitment in the O-phase.

In contrast to Mayer's strategy, Hardy-Kent's strategy is symmetric with
respect to the value of $b$ that Alice unveils in the O-phase. The lower
limit $P_{HK\text{ }err}^{Alice}(\rho _{0}^{B},\rho _{1}^{B})$ is
independent of her disclosure of $b$. \ The upper bound of the mutual
information $I_{HK}^{Alice}$ for Hardy-Kent's strategy is written in terms
of $P_{HK\text{ }err}^{Alice}(\rho _{0}^{B},\rho _{1}^{B})$ as 
\begin{equation}
I_{HK}^{Alice}(\rho _{0}^{B},\rho _{1}^{B})=1-H(P_{HK\text{ }%
err}^{Alice}(\rho _{0}^{B},\rho _{1}^{B})),  \label{eq28}
\end{equation}%
which is considered to be a good information-theoretic measure of Alice's
ability to change commit bit $b$ in this strategy.

\begin{figure}[tbp]
\epsfxsize=8.5cm \epsfbox{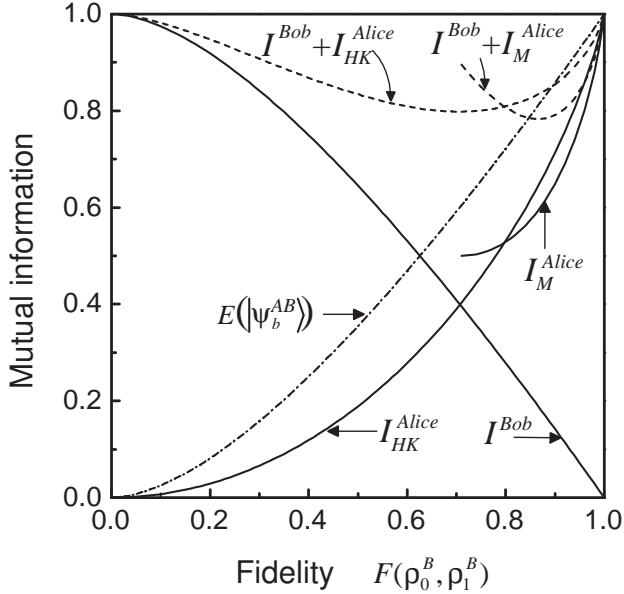}
\caption{\hspace{2mm} Three \hspace{6mm} information-theoretic \hspace{6mm} measures \newline
$I^{Bob}(\rho _{0}^{B},\rho _{1}^{B})$, $I_{M}^{Alice}(\rho _{0}^{B},%
\rho _{1}^{B})$, and $I_{HK}^{Alice}(\rho _{0}^{B},%
\rho _{1}^{B})$ are plotted against fidelity $F(\rho _{0}^{B},%
\rho _{1}^{B})$. Entropy of entanglement $E\left( \left| %
\psi _{b}^{AB}\right\rangle \right) $ is also plotted for reader's information.}
\label{F1}
\end{figure}

To compare the cheating performances of Alice and Bob for both Mayer's and
Hardy-Kent's strategies, we plot the three information theoretic measures $%
I^{Bob}(\rho _{0}^{B},\rho _{1}^{B})$, $I_{M}^{Alice}(\rho _{0}^{B},\rho
_{1}^{B})$, and $I_{HK}^{Alice}(\rho _{0}^{B},\rho _{1}^{B})$ in Fig. \ref%
{F1} as a function of the fidelity $F(\rho _{0}^{B},\rho _{1}^{B})$ chosen
as a common parameter. This figure clearly shows that there is a trade-off
relationship between Bob's available information in the C-phase ($%
I^{Bob}(\rho _{0}^{B},\rho _{1}^{B})$) and Alice's ability to change commit
bit $b$ in the O-phase ($I_{i}^{Alice}(\rho _{0}^{B},\rho _{1}^{B})$). It is
clear that the sum is bounded; i.e., 
\begin{equation}
I^{Bob}(\rho _{0}^{B},\rho _{1}^{B})+I_{i}^{Alice}(\rho _{0}^{B},\rho
_{1}^{B})\leq 1  \label{eq29}
\end{equation}%
(for $i=M,HK$). This equation is a direct consequence of Eq. (\ref{eq15}),
showing a trade-off relationship between the distinguishability $D(\rho
_{0}^{B},\rho _{1}^{B})$, a measure of Bob's information gain in the
C-phase, and the fidelity $F(\rho _{0}^{B},\rho _{1}^{B})$, a measure of
Alice's ability to change commit bit in the O-phase. Therefore, there is a
trade-off in the performance of Alice's and Bob's cheating. Figure \ref{F1}
also shows that Hardy-Kent's strategy is superior to Mayer's with respect to
ability to change commit bit $b$ when the value of $F(\rho _{0}^{B},\rho
_{1}^{B})$ is large. 

\section{Discussion}

\label{SEC4}

A secure BC protocol must not allow cheating by either parties, Alice or
Bob. To satisfy this condition, both $I^{Bob}(\rho _{0}^{B},\rho _{1}^{B})$
and $I_{i}^{Alice}(\rho _{0}^{B},\rho _{1}^{B})$ should vanish
simultaneously. However, Fig. \ref{F1} indicates that this requirement is
never satisfied because of the trade-off relationship between $I^{Bob}(\rho
_{0}^{B},\rho _{1}^{B})$ and $I_{i}^{Alice}(\rho _{0}^{B},\rho _{1}^{B})$.
In addition, even if we choose a balanced condition for both parties, $%
F(\rho _{0}^{B},\rho _{1}^{B})\sim 1/\sqrt{2}$, both $I^{Bob}(\rho
_{0}^{B},\rho _{1}^{B})$ and $I_{i}^{Alice}(\rho _{0}^{B},\rho _{1}^{B})$
are already large enough. Therefore, it is concluded that a law of quantum
physics does not help to improve the security of the BC protocol.

As already proven generally by Mayers, Lo, and Chau, this conclusion should
be valid not only for the particular cheating strategies described in this
paper but also for any strategies that Alice and Bob can choose. To
understand this conclusion, consider the entropy of entanglement (or
entanglement in brief), which is known to be a unique measure of the amount
of entanglement for the pure state.\cite{Barnett,Bennett2,Plenio}
Entanglement of the purification $\left| \psi _{b}^{AB}\right\rangle $ is
defined as the von Neumann entropy of the marginal state $\rho _{b}^{B}$ of $%
\left| \psi _{b}^{AB}\right\rangle $ or equivalently as the Shannon entropy
of the squares of the Schmidt coefficients of $\left| \psi
_{b}^{AB}\right\rangle $. From Eqs. (\ref{eq4}), (\ref{eq5}), (\ref{eq13}),
and (\ref{eq16}), it is easily calculated as 
\begin{equation}
E\left( \left| \psi _{b}^{AB}\right\rangle \right) =S\left( \rho
_{b}^{B}\right) =1-I^{Bob}(\rho _{0}^{B},\rho _{1}^{B}).  \label{eq30}
\end{equation}%
Here, it should be noted that $I^{Bob}(\rho _{0}^{B},\rho _{1}^{B})$ is
equivalent to maximum information $I_{c}$ available from $\rho _{b}^{B}$
that is transmitted from Alice to Bob in the C-phase; that is, 
\begin{equation}
I_{c}=I^{Bob}(\rho _{0}^{B},\rho _{1}^{B})=1-E\left( \left| \psi
_{b}^{AB}\right\rangle \right) .  \label{eq31}
\end{equation}%
Applying Eqs. (\ref{eq1}) and (\ref{eq29}) to (\ref{eq31}), we obtain the
inequality,  
\begin{equation}
I_{i}^{Alice}(\rho _{0}^{B},\rho _{1}^{B})\leq E\left( \left| \psi
_{b}^{AB}\right\rangle \right) =I_{o}.  \label{eq32}
\end{equation}%
This inequality implies two things. First, the performance of Alice's
cheating, when it is measured by $I_{i}^{Alice}(\rho _{0}^{B},\rho _{1}^{B})$%
, is bounded by the entanglement of the purification $E\left( \left| \psi
_{b}^{AB}\right\rangle \right) $, which is determined only by its marginal
state $\rho _{b}^{B}$ (see Eq. (\ref{eq30})). Second, it is also bounded by
the amount of information $I_{o}$ that is revealed in the O-phase.

These implications are reasonable for the following reason. When Alice wants
to cheat Bob, what she can do is restricted to local operation and
measurement on the subsystem $A$ in her hand after she has revealed $\rho
_{b}^{B}$ by transmitting the subsystem $B$. It is known to be a fundamental
law of quantum information processing that entanglement cannot be increased
if we are allowed to perform only local operations and subselection on the
subsystem of a joint system. In this restricted situation, the best she can
do to cheat is use the local unitary operation that conserves the
entanglement shared in the joint system and keeps the marginal states $\rho
_{b}^{B}$ unchanged. Otherwise, the strategy must be by far an optimal one
because a fraction of the entanglement must be lost from the joint system
and dissipate into the environment during the local operation. Under such
circumstances, Alice can change the information content encoded only in the
relative phase between coefficients of each term in the purification $\left|
\psi _{b}^{AB}\right\rangle $, but she cannot change the information content
encoded in their absolute values. It is the entanglement resource that is
responsible for Alice's cheating, and there is no cheating strategy that can
break the bound given by entanglement $E\left( \left| \psi
_{b}^{AB}\right\rangle \right) $ as shown in Eq. (\ref{eq32}). In addition,
it is also reasonable that only partial information that is to be revealed
in the O-phase can be used for Alice's cheating but the partial information
already revealed in the C-phase cannot. Conversely, we must be aware that
Alice makes use of partial information that is reserved to be revealed in
the O-phase as an entanglement resource for cheating.

It is worth noting that the present proof can be regarded as a concrete
example of the general proof of the no-go theorem for a
zero-knowledge-convincing protocol recently given by Horodecki et al.\cite%
{Horodecki} Our proof clearly indicates that if Alice wants to convince Bob
that she has a definite value of a commit bit (which is, of course, {\it %
classical} information) in mind in the C-phase, the information provided by
her to him in the C-phase has to carry nontrivial information about the
commit bit in her mind. If the information revealed in the C-phase is
independent of her commit bit, Alice can always try to cheat by proposing
the test which would give some result with certainty and independently of
her commit bit in the O-phase. Our proof suggests an information-theoretic
ground for the no-go theorem of the zero-knowledge-convincing protocol.
Namely, any protocol with a test message that convinces Bob that Alice knows
some state $\phi $ (which is, in general, {\it quantum} information), the
test message has to carry non-zero information about state $\phi $ to
prevent Alice's cheating.

The present proof implies that the conjecture of Mayers about two-party
secure computation, which states that the symmetric protocol might be
possible whereas the asymmetric tasks, such as unidirectional secure
computations, would be impossible, is correct.\cite{Mayers} In the
unidirectional two-party computation, which allows only one of the two
parties to learn the result, both members of the party can be a cheater and
security requirements for both members are incompatible. Such unidirectional
protocols under the standard nonrelativistic assumption are necessarily
insecure. We believe that unidirectional quantum communication does not
achieve more than classical communication alone in the two-party model.
However, it has not yet been proven that no non-trivial cryptographic tasks
in the two-party model using bidirectional quantum communication are
unconditionally secure. Indeed, there are some proposals on the quantum
protocols for non-trivial weaker tasks in two-party bidirectional quantum
communication such as quantum coin-tossing\cite{Mayers2} and quantum
gambling.\cite{Goldenberg} It will still be important to solve the general
problem concerning what is possible and what is impossible in two-party
secure computation when unproven computational assumptions are abandoned.

\section{Conclusions}

\label{SEC5}

In conclusion, we have given constructive proof why an unconditionally
secure quantum BC is impossible in the light of quantum information theory.
The BC protocol is in essence the protocol in which one-bit information is
split and revealed in two temporally separated steps: the C-phase and the
O-phase. It ensures only a fraction of the bit information is revealed at a
time. In the quantum BC protocol, increasing the information revealed in the
C-phase is to Bob's advantage; conversely, increasing the information
revealed in the O-phase makes things to Alice's advantage. This situation is
similar to the classical protocol. Furthermore, the protocol that is
unbiased to both Alice and Bob is not secure for both. Therefore, it is
impossible to design a BC protocol whose security is established solely on
the law of quantum physics.

In addition, it has been clarified that, Alice can make use of the
entanglement resource, which is equal to the amount of information reserved
to be revealed in the O-phase, to cheat. To prevent Alice's cheating, the
information revealed in the C-phase must depend on her commit bit, and it
must inevitably carry non-zero information about her commitment. It can be
concluded that quantum mechanics itself makes designing an unconditionally
secure BC protocol impossible.

\appendix

\section{A proof of existence of mutually orthogonal purifications of $%
\protect\rho _{\lowercase{b}}^{B}$}

\label{APPA}

Suppose that the states $\chi _{b}^{AB}$ ($b=0,1$) of joint system $AB$ that
is to be prepared by Alice are mutually orthogonal on the joint space $%
H_{AB}=H_{A}\otimes H_{B}$; i.e., 
\begin{equation}
\chi _{0}^{AB}\chi _{1}^{AB}=\chi _{1}^{AB}\chi _{0}^{AB}=0.  \label{A1}
\end{equation}%
Because $\chi _{0}^{AB}$ and $\chi _{1}^{AB}$ commute, they can be
diagonalized simultaneously in terms of orthonormal bases $\left\{ \left|
e^{AB}\right\rangle \right\} $ and $\left\{ \left| f^{AB}\right\rangle
\right\} $ in $H_{AB}$ as follows: 
\begin{eqnarray}
\chi _{0}^{AB} &=&\sum \lambda _{e}\left| e^{AB}\right\rangle \left\langle
e^{AB}\right| ,  \nonumber \\
\chi _{1}^{AB} &=&\sum \lambda _{f}\left| f^{AB}\right\rangle \left\langle
f^{AB}\right| ,  \label{A2}
\end{eqnarray}%
where $\left\{ \left| e^{AB}\right\rangle \right\} $ and $\left\{ \left|
f^{AB}\right\rangle \right\} $ are mutually orthogonal; i.e., $\left\langle
e^{AB}|f^{AB}\right\rangle =\left\langle f^{AB}|e^{AB}\right\rangle =0$, and 
$\left\{ \lambda _{e}\right\} $ and $\left\{ \lambda _{f}\right\} $ are sets
of real eigenvalues of $\chi _{b}^{AB}$ satisfying $0\leq \lambda
_{e},\lambda _{f}\leq 1$ and $\sum \lambda _{e}=\sum \lambda _{f}=1$. Thus, $%
\chi _{0}^{AB}$ and $\chi _{1}^{AB}$ have orthogonal supports in $H_{AB}$.
Marginal states revealed by Alice to Bob in the C-phase are commutable and,
in general, nonorthogonal states. Using this representation, we can write
them as 
\begin{eqnarray}
\rho _{0}^{B} &=&Tr_{A}\chi _{0}^{AB}=\sum \lambda _{e}Tr_{A}\left|
e^{AB}\right\rangle \left\langle e^{AB}\right| ,  \nonumber \\
\rho _{1}^{B} &=&Tr_{A}\chi _{1}^{AB}=\sum \lambda _{f}Tr_{A}\left|
f^{AB}\right\rangle \left\langle f^{AB}\right| .  \label{A3}
\end{eqnarray}%
Now, we consider mutually orthonormal states $\left| \psi
_{0}^{AB}\right\rangle $ and $\left| \psi _{1}^{AB}\right\rangle $ $%
(\left\langle \psi _{0}^{AB}|\psi _{1}^{AB}\right\rangle =0)$ that lie in
the subspace to which $\chi _{0}^{AB}$ and $\chi _{1}^{AB}$ belong
respectively; i.e., 
\begin{eqnarray}
\left| \psi _{0}^{AB}\right\rangle  &=&\sum c_{e}\left| e^{AB}\right\rangle ,
\nonumber \\
\left| \psi _{1}^{AB}\right\rangle  &=&\sum c_{f}\left| f^{AB}\right\rangle .
\label{A4}
\end{eqnarray}%
Then, the marginal states for them are 
\begin{eqnarray}
Tr_{A}\left| \psi _{0}^{AB}\right\rangle \left\langle \psi _{0}^{AB}\right| 
&=&\sum \left| c_{e}\right| ^{2}Tr_{A}\left| e^{AB}\right\rangle
\left\langle e^{AB}\right|   \nonumber \\
&+&\sum c_{e}c_{e^{\prime }}^{\ast }Tr_{A}\left| e^{AB}\right\rangle
\left\langle e^{\prime AB}\right| ,  \nonumber \\
Tr_{A}\left| \psi _{1}^{AB}\right\rangle \left\langle \psi _{1}^{AB}\right| 
&=&\sum \left| c_{f}\right| ^{2}Tr_{A}\left| f^{AB}\right\rangle
\left\langle f^{AB}\right|   \nonumber \\
&+&\sum c_{f}c_{f^{\prime }}^{\ast }Tr_{A}\left| f^{AB}\right\rangle
\left\langle f^{\prime AB}\right| .  \label{A5}
\end{eqnarray}%
Because the states of subsystem $A$ represent the classical information
transferred from Alice to Bob in the O-phase, different states $\left|
e^{AB}\right\rangle \neq \left| e^{\prime AB}\right\rangle $ and $\left|
f^{AB}\right\rangle \neq \left| f^{\prime AB}\right\rangle $ are orthogonal
on the subspace $H_{A}$. Therefore, 
\begin{equation}
Tr_{A}\left| e^{AB}\right\rangle \left\langle e^{\prime AB}\right|
=Tr_{A}\left| f^{AB}\right\rangle \left\langle f^{\prime AB}\right| =0.
\label{A6}
\end{equation}%
By noting that the second terms in Eq. (\ref{A5}) vanishes, it is concluded
that by choosing $c_{e}$ and $c_{f}$ so that 
\begin{eqnarray}
\lambda _{e} &=&\left| c_{e}\right| ^{2},  \nonumber \\
\lambda _{f} &=&\left| c_{f}\right| ^{2},  \label{A7}
\end{eqnarray}%
it is always possible to obtain mutually orthogonal purification $\left|
\psi _{b}^{AB}\right\rangle $\ of $\rho _{b}^{AB}$ in the subspace in which
the support of the state $\chi _{b}^{AB}$ lies.

\section{Probability that Bob detects Alice's cheating}

\label{APPB}

Suppose that the state of subsystem $A$ is measured to be $\left|
e_{j}^{A}\right\rangle $ when $A$\ of the joint system $AB$ prepared in the
state $\left| \psi _{b}^{AB}\right\rangle \left\langle \psi _{b}^{AB}\right| 
$ is subjected to projection measurement by the orthonormal basis $\left\{
\left| e_{j}^{A}\right\rangle \right\} $ for $H_{A}$. According to general
results of quantum measurement theory, the state of the subsystem $B$ is
projected onto the pure state 
\begin{equation}
\rho _{b}^{B}\left( \left| e_{j}^{A}\right\rangle \right) =\frac{%
\left\langle e_{j}^{A}|\psi _{b}^{AB}\right\rangle \left\langle \psi
_{b}^{AB}|e_{j}^{A}\right\rangle }{Tr_{B}\left\langle e_{j}^{A}|\psi
_{b}^{AB}\right\rangle \left\langle \psi _{b}^{AB}|e_{j}^{A}\right\rangle }.
\label{B1}
\end{equation}%
From Eqs. (\ref{eq4}) and (\ref{eq5}), it is easily seen that 
\begin{eqnarray}
\left\langle \psi _{0}^{AB}|e_{j}^{A}\right\rangle \left\langle
e_{j}^{A}|\psi _{1}^{AB}\right\rangle  &=&\frac{\sin 2\theta }{2}\left(
\left\langle 1^{AB}|e_{j}^{A}\right\rangle \left\langle
e_{j}^{A}|1^{AB}\right\rangle \right.   \nonumber \\
&&-\left. \left\langle 0^{AB}|e_{j}^{A}\right\rangle \left\langle
e_{j}^{A}|0^{AB}\right\rangle \right) .  \label{B2}
\end{eqnarray}%
Therefore, we find that, if and only if the basis $\left\{ \left|
e_{j}^{A}\right\rangle \right\} $ is chosen so that the overlap between $%
\left| e_{j}^{A}\right\rangle $ and $\left| 0^{AB}\right\rangle $ and that
between $\left| e_{j}^{A}\right\rangle $ and $\left| 1^{AB}\right\rangle $
are the same, i.e., 
\begin{equation}
\left\langle 0^{AB}|e_{j}^{A}\right\rangle \left\langle
e_{j}^{A}|0^{AB}\right\rangle =\left\langle 1^{AB}|e_{j}^{A}\right\rangle
\left\langle e_{j}^{A}|1^{AB}\right\rangle ,  \label{B3}
\end{equation}%
the states $\rho _{0}^{B}\left( \left| e_{j}^{A}\right\rangle \right) $ and $%
\rho _{1}^{B}\left( \left| e_{j}^{A}\right\rangle \right) $ become mutually
orthogonal. In the quantum BC protocol, Alice and Bob agree to use the
measurement basis $\left\{ \rho _{0}^{B}\left( \left| e_{j}^{A}\right\rangle
\right) ,\rho _{1}^{B}\left( \left| e_{j}^{A}\right\rangle \right) \right\} $
on subsystem $B$\ that has a one-to-one correspondence to a state $\left|
e_{j}^{A}\right\rangle $ on $A$ through the joint state $\left| \psi
_{b}^{AB}\right\rangle \left\langle \psi _{b}^{AB}\right| $. She reveals to
Bob the measurement basis $\left\{ \rho _{0}^{B}\left( \left|
e_{j}^{A}\right\rangle \right) ,\rho _{1}^{B}\left( \left|
e_{j}^{A}\right\rangle \right) \right\} $ associated with her state $\left|
e_{j}^{A}\right\rangle $ in the O-phase, and he measures his subsystem $B$
by this basis.

Now, we consider the probability $P_{err}^{Alice}$\ that Alice causes an
inconsistency between the value of $b$ that is unveiled by her in the
O-phase and Bob's measured data when Alice prepares an honest state $\chi
_{b}^{AB}$. Alice projects her subsystem $A$ of the joint system $AB$
prepared in $\chi _{b}^{AB}$ onto a state $\left| e_{j}^{A}\right\rangle $
among the complete orthonormal basis $\left\{ \left| e_{j}^{A}\right\rangle
\right\} $ for space $H_{A}$. She can perform such a projection on her
subsystem at her own free will. Correspondingly, the state of Bob's system
is projected to be 
\begin{equation}
\tilde{\rho}_{b}^{B}\left( \left| e_{j}^{A}\right\rangle \right) =\frac{%
\left\langle e_{j}^{A}\right| \chi _{b}^{AB}\left| e_{j}^{A}\right\rangle }{%
Tr_{B}\left\langle e_{j}^{A}\right| \chi _{b}^{AB}\left|
e_{j}^{A}\right\rangle }.  \label{B4}
\end{equation}%
Here, from Appendix A, the states $\left| \psi _{b}^{AB}\right\rangle
\left\langle \psi _{b}^{AB}\right| $ and $\chi _{b}^{AB}$ satisfy 
\begin{equation}
\left\langle e_{j}^{A}|\psi _{b}^{AB}\right\rangle \left\langle \psi
_{b}^{AB}|e_{j}^{A}\right\rangle =\left\langle e_{j}^{A}\right| \chi
_{b}^{AB}\left| e_{j}^{A}\right\rangle .  \label{B5}
\end{equation}%
Then, we obtain the identity $\rho _{b}\left( \left| e_{j}^{A}\right\rangle
\right) =\tilde{\rho}_{b}\left( \left| e_{j}^{A}\right\rangle \right) $.
This identity implies that if Bob follows Alice's instruction and measures
his system by the measurement basis given by her, the value of $b$ unveiled
by Alice in the O-phase is perfectly correlated with Bob's measurement
result, no matter what Alice prepares $\left| \psi _{b}^{AB}\right\rangle
\left\langle \psi _{b}^{AB}\right| $ or $\chi _{b}^{AB}$. Therefore, if Bob
is honest enough to follow Alice's instruction, the probability $%
P_{err}^{Alice}$\ that Bob finds an inconsistency in his data vanishes if
Alice prepares $\left| \psi _{b}^{AB}\right\rangle \left\langle \psi
_{b}^{AB}\right| $ or $\chi _{b}^{AB}$. Consequently, Alice can transmit one
bit of classical information to Bob with certainty.

Next, we consider the probability $P_{err}^{Alice}$ when Alice prepares a
fake state $\left| \bar{\psi}^{AB}\right\rangle $ that lies in joint space $%
H_{AB}$. When the joint system $AB$ prepared in the state $\left| \bar{\psi}%
^{AB}\right\rangle \left\langle \bar{\psi}^{AB}\right| $ is subjected to the
projection measurement by using the orthonormal basis $\left\{ \left|
e_{j}^{A}\right\rangle \right\} $ for $H_{A}$ and the result is $\left|
e_{j}^{A}\right\rangle $, the state of the subsystem $B$ is projected onto
the pure state 
\begin{equation}
\bar{\rho}^{B}\left( \left| e_{j}^{A}\right\rangle \right) =\frac{%
\left\langle e_{j}^{A}|\bar{\psi}^{AB}\right\rangle \left\langle \bar{\psi}%
^{AB}|e_{j}^{A}\right\rangle }{Tr_{B}\left\langle e_{j}^{A}|\bar{\psi}%
^{AB}\right\rangle \left\langle \bar{\psi}^{AB}|e_{j}^{A}\right\rangle }.
\label{B6}
\end{equation}%
Let the fidelity between $\bar{\rho}^{B}\left( \left| e_{j}^{A}\right\rangle
\right) $ and $\rho _{b}^{B}\left( \left| e_{j}^{A}\right\rangle \right) $
be $F\left( \bar{\rho}^{B}\left( \left| e_{j}^{A}\right\rangle \right) ,\rho
_{b}^{B}\left( \left| e_{j}^{A}\right\rangle \right) \right) $. Then, the
probability $P_{err}^{Alice}$\ that Bob finds an inconsistency in his data
is given by 
\begin{equation}
P_{err}^{Alice}=1-\left| F\left( \bar{\rho}^{B}\left( \left|
e_{j}^{A}\right\rangle \right) ,\rho _{b}^{B}\left( \left|
e_{j}^{A}\right\rangle \right) \right) \right| ^{2}.  \label{B7}
\end{equation}%
Under the condition of Eq. (\ref{B3}), it follows that 
\begin{eqnarray}
\left\langle \psi ^{AB}|e_{j}^{A}\right\rangle \left\langle e_{j}^{A}|\psi
_{b}^{AB}\right\rangle =\frac{1}{2}Tr_{AB} &&\left| e_{j}^{A}\right\rangle
\left\langle e_{j}^{A}\right| P_{M}  \nonumber \\
&&\cdot \left\langle \psi ^{AB}|\psi _{b}^{AB}\right\rangle ,  \label{B8}
\end{eqnarray}%
%
%
%
%
%
\begin{eqnarray}
Tr_{B}\left\langle e_{j}^{A}|\psi ^{AB}\right\rangle \left\langle \psi
^{AB}|e_{j}^{A}\right\rangle  &=&Tr_{B}\left\langle e_{j}^{A}|\psi
_{b}^{AB}\right\rangle \left\langle \psi _{b}^{AB}|e_{j}^{A}\right\rangle  
\nonumber \\
&\geq &\frac{1}{2}Tr_{AB}\left| e_{j}^{A}\right\rangle \left\langle
e_{j}^{A}\right| P_{M},  \label{B9}
\end{eqnarray}%
where $P_{M}=\left| 0^{AB}\right\rangle \left\langle 0^{AB}\right| +\left|
1^{AB}\right\rangle \left\langle 1^{AB}\right| $ is the projector onto
two-dimensional subspace $M$ in $H_{AB}$ that is spanned by a set of
orthonormal states $\left\{ \left| 0^{AB}\right\rangle ,\left|
1^{AB}\right\rangle \right\} $, and $Tr_{AB}\left| e_{j}^{A}\right\rangle
\left\langle e_{j}^{A}\right| P_{M}=\left\langle
0^{AB}|e_{j}^{A}\right\rangle \left\langle e_{j}^{A}|0^{AB}\right\rangle
+\left\langle 1^{AB}|e_{j}^{A}\right\rangle \left\langle
e_{j}^{A}|1^{AB}\right\rangle $ is the overlap between the state $\left|
e_{j}^{A}\right\rangle $ in $H_{A}$ and subspace $M$. The equal sign in
inequality (\ref{B9}) holds if and only if state $\left| \bar{\psi}%
^{AB}\right\rangle $ lies within subspace $M$.

From Eqs. (\ref{B1}),(\ref{B6}),(\ref{B8}), and (\ref{B9}), we obtain 
\begin{equation}
\left| F\left( \bar{\rho}^{B}\left( \left| e_{j}^{A}\right\rangle \right)
,\rho _{b}^{B}\left( \left| e_{j}^{A}\right\rangle \right) \right) \right|
^{2}\leq \left| \left\langle \psi ^{AB}|\psi _{b}^{AB}\right\rangle \right|
^{2}.  \label{B10}
\end{equation}%
Applying Eq. (\ref{B10}) to Eq. (\ref{B7}), we finally obtain 
\begin{equation}
P_{err}^{Alice}\geq 1-\left| \left\langle \psi ^{AB}|\psi
_{b}^{AB}\right\rangle \right| ^{2}.  \label{B11}
\end{equation}%
Here, equality holds if and only if the state $\left| \bar{\psi}%
^{AB}\right\rangle $ lies within subspace $M$.


\end{document}